# Amplification of valley polarization in an hBN-separated $MoSe_2$-$WSe_2$ heterostructure

Fateme Mahdikhany[1,2†], Jacob Cutshall[1†], Daniel Shanks[1], Garett Reichenbach[1], Davoud Adinehloo[3], Michael R. Koehler[4], David G. Mandrus[5-7], Takashi Taniguchi[8], Kenji Watanabe[9], Vasili Perebeinos[3], Brian J. LeRoy[1], John R. Schaibley[1]

[1]Department of Physics, University of Arizona, Tucson, Arizona 85721, USA

[2]Department of Materials Science and Engineering, Northwestern University, Evanston, Illinois 60208, USA

[3]Department of Electrical Engineering, University at Buffalo, Buffalo, New York, 14260

[4]IAMM Diffraction Facility, Institute for Advanced Materials and Manufacturing, University of Tennessee, Knoxville, TN 37920

[5]Department of Materials Science and Engineering, University of Tennessee, Knoxville, Tennessee 37996, USA

[6]Materials Science and Technology Division, Oak Ridge National Laboratory, Oak Ridge, Tennessee 37831, USA

[7]Department of Physics and Astronomy, University of Tennessee, Knoxville, Tennessee 37996, USA

[8]Research Center for Materials Nanoarchitectonics, National Institute for Materials Science, 1-1 Namiki, Tsukuba 305-0044, Japan

[9]Research Center for Electronic and Optical Materials, National Institute for Materials Science, 1-1 Namiki, Tsukuba 305-0044, Japan

[†]Fateme Mahdikhany and Jacob Cutshall are equal contributors to this work and designated as co-first authors.
Corresponding Author: John Schaibley, johnschaibley@arizona.edu

**ABSTRACT**. Transition metal dichalcogenide heterostructures host long-lived interlayer excitons in the +K and -K valleys that can encode long-lived valley polarization. We investigated the low-temperature valley polarization dynamics in an $MoSe_2$-bilayer hBN-$WSe_2$ heterostructure. The bilayer hBN spacer results in an interlayer exciton lifetime exceeding 2 μs and an anomalous valley polarization amplification effect. Our theoretical model shows that this amplification results from the long exciton lifetime and exciton-exciton interactions. This work sets the stage for long-lived interlayer excitons to be used in nonlinear valleytronic transistors and amplifiers.

## I. INTRODUCTION.

Excitons in transition metal dichalcogenide (TMD) heterostructures can be selectively excited with circularly polarized light to either the +K or -K valley and can encode information of which valley they occupy [1-6]. These excitons have been shown be the foundation of tunable quantum dots [7,8], correlated states [9,10], 2D optoelectronics [11], and excitonic valleytronics [1-4,12-15]. One key restriction in these systems is the valley polarization lifetime: the time before a valley-polarized exciton scatters from one valley to the other. In monolayer TMDs, excitonic valley depolarization typically occurs within picoseconds [15-18]. Direct contact (DC) $MoSe_2/WSe_2$ heterostructures lead to the formation of unique Moiré physics [19-21], and the creation of an interlayer exciton (IX) leads to further suppression of this scattering [22-24], resulting in valley depolarization time of approximately ~10-100 ns [25]. In this work, we perform time and valley polarization resolved low temperature photoluminescence (PL) measurements of interlayer excitons in bilayer hBN separated $MoSe_2$-$WSe_2$ heterostructures. We show that, unlike in DC heterostructures and in other ungated monolayer hBN separated devices [25], the degree of valley polarization in hBN-separated regions increases in time. We note that while there are numerous studies of IXs in $MoSe_2/WSe_2$ heterostructures, we report a highly anomalous valley polarization amplification effect in the hBN separated region.

## II. METHODS

The 2D material layers were individually isolated from the bulk using the Scotch tape method and assembled with a 1-degree relative twist angle using a polymer based dry transfer [26]. The sample (Supplementary Figure 1 [27]) is composed of a bottom hBN layer (22 nm), an 8 nm top hBN layer, top bilayer graphene, and a 2 nm back gate of graphite. We used electron beam lithography and thermal evaporation to fabricate chromium-gold contacts on the device.

The sample was investigated using low-temperature PL spectroscopy. The sample was held at 1.6 K in an optical cryostat (attocube attodry 2100) with an 0.82 NA objection. For all the measurements presented, we used 723 nm (1.715 eV) excitation wavelength, near resonance with the $WSe_2$ exciton. Lifetime measurements were taken by exciting with 120 fs pulses, separated by 8000 ns. 8000 ns separation was achieved by pulse picking the output of a tunable femtosecond Ti:sapphire laser using an acousto-optic modulator, or a wavelength filtered supercontinuum laser with a pulse width of ~100 ps and a tunable repetition rate. To independently modulate the electric field and doping density, we applied voltages to the gate in a ratio matching the ratio of the top to bottom hBN thickness. A confocal pinhole with an effective collection diameter of 1.4 μm was used to spatially resolve the PL emission. Lifetime measurements were conducted using a time-correlated single photon counting system and a single photon avalanche detector.

## III. RESULTS

A schematic of the hBN-encapsulated $MoSe_2$-$WSe_2$ heterostructure is shown in Fig. 1(a). This device has two distinct regions: a bilayer hBN-separated region and a direct contact (DC) region where $MoSe_2$ and $WSe_2$ are directly contacting each other. Fig. 1(b) shows a confocal PL map of the IX emission as a function of position.

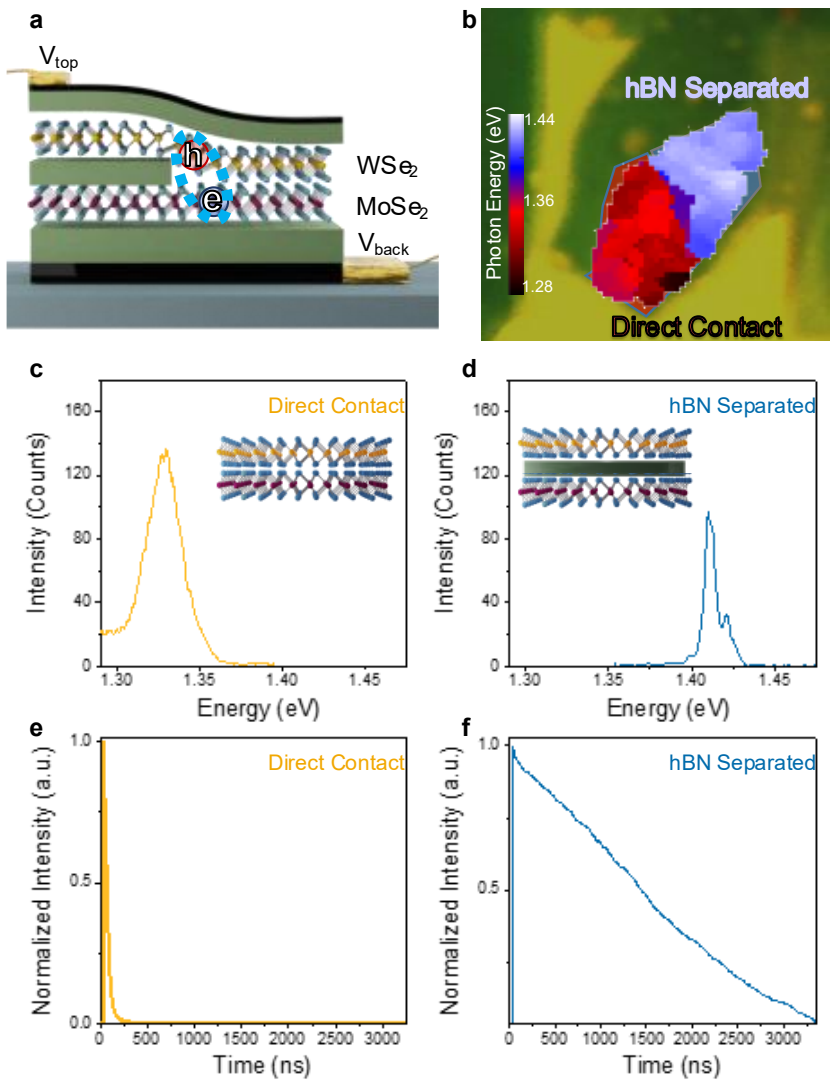

**FIG 1. hBN-separated and direct contact IX PL and lifetime.** (a) Illustration of the device. (b) A spatial map of the PL energy center. (c) An example PL from the DC region. (d) An example PL from the hBN-separated region. (e) Time-dependent PL shows a lifetime of ~50 ns for the direct contact and ~2000 ns for the hBN-separated area (f).

As reported previously, the hBN-separated IXs (with near zero twist angle alignment) emit PL at 1.42 eV [28], while the DC region continues to emit PL centered at 1.33 eV, consistent with previous reports of R-type heterostructures [4,29-31]. Fig. 1(c) and 1(d) display example PL spectra from the two distinct areas. Time-dependent PL measurements are shown in Fig. 1(e). The DC region exhibits a lifetime of ~40 ns. In contrast, the hBN-separated area has an extended lifetime of ~2000 ns (Fig. 1(f)). This lifetime can reach approximately 2800 ns by increasing excitation power, (Supplementary Fig. 2). This increase in lifetime is attributed to the reduced spatial overlap of the IX's electron and hole wave functions.

We also investigated the effect of electron and hole doping on the IX's lifetime. The system is doped by simultaneously varying the top and bottom gates, with no electric field applied. The resulting steady-state PL as a function of doping within the hBN-separated region is shown in Fig. 2(a). We observe that hole doping the system results in the appearance of the positive IX trion peak, leading to a brighter signal. In contrast, electron doping causes a reduction in the signal intensity within the hBN-separated region.

Time resolved PL data was fit with a double exponential for each doping level. The doping dependence of the longer lifetime component is plotted in Fig. 2(b), and the shorter lifetime component (which does not have a strong doping dependence) is shown in Supplementary Fig. 3. The data derived from these measurements shows a neutral IX lifetime of roughly 1000 ns. When the sample is hole doped, the lifetime is nearly constant around 1800 ns. Alternatively, electron doping initially increases the lifetime to near 1600 ns, which then drops to approximately 600 ns at 0 doping.

This observed behavior deviates from previous reports in similar DC devices [17,18,22]. Steady state doping dependent PL measurements on the DC region are shown in Fig. 2(c). Both electron and hole doping give rise to IX trion peaks that appear at lower energies than the neutral peak. Fig. 2(d) shows the results of the doping dependent lifetime data. Simultaneously, as the trion peaks turn on, the IX lifetime decreases from 45.89 ns to 40.13 ns, as presented in Fig. 2(d). This finding is consistent with prior measurements in analogous systems [32]. This reveals a non-trivial trend with doping in the hBN-separated region. Electric field dependent lifetime of both regions are shown in Supplementary Fig. 4.

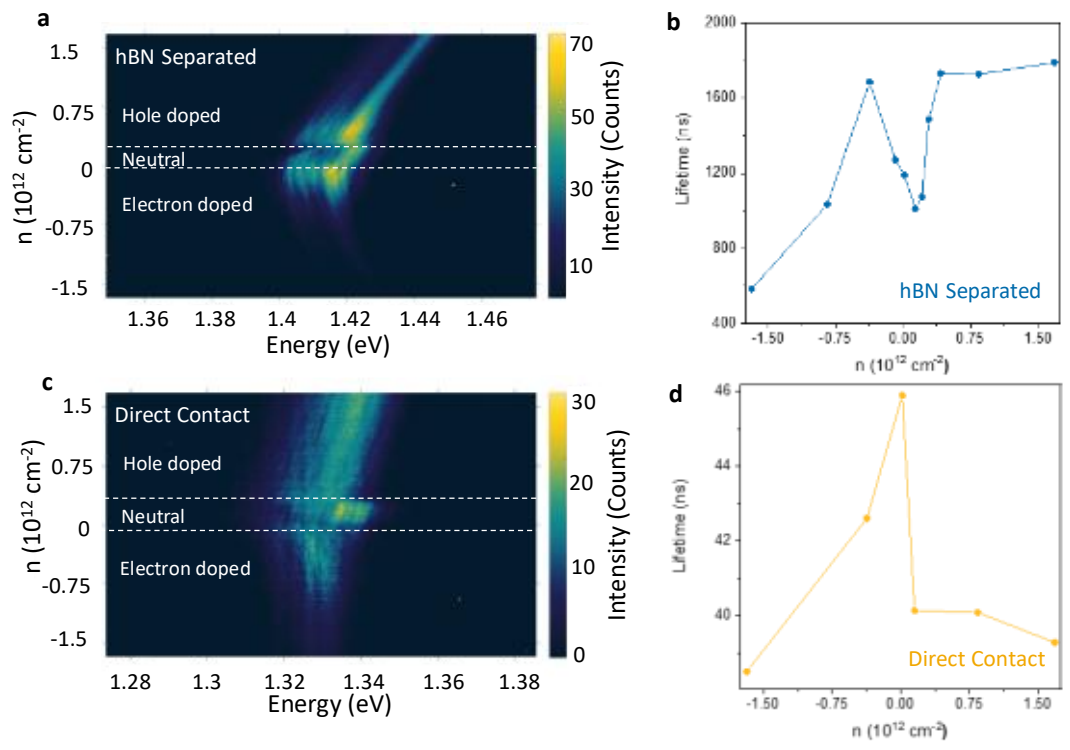

**FIG 2. PL and lifetime as function of carrier density.** (a) PL as a function of doping density in the hBN separated area and the direct contact area (c). (b) Lifetime of the IX as a function of doping density for the hBN separated area and direct contact area (d).

The co- and cross- circularly polarized components of the time resolved PL was measured in the hBN separated region by exciting with right-hand polarized (RHP) light and detecting both RHP and left-hand polarized (LHP) PL. The degree of polarization, $P = \frac{I_{\sigma+} - I_{\sigma-}}{I_{\sigma+} + I_{\sigma-}}$, as a function of time is shown in Fig. 3(a). $I_{\sigma\pm}$ are the intensities of the LHP and RHP components. Notably, the data indicates that, when the sample is neutral, the degree of polarization amplifies over time.

As depicted in Fig. 3(b), for a neutral sample, the intensity of the cross-polarized component of the lifetime decreases more rapidly than that of the co-polarized component. This differential decay leads to an enhancement in polarization (blue curve in Fig. 3(b)), achieving 75% polarization above 3000 ns. The result of a similar measurement in the DC area is shown in Supplementary Fig. 5. As expected, we observed a decay of the valley polarization over time, which is consistent with previous findings [22-24].

Figure 3(c) shows time dependent polarization as a function of excitation power, showing that the initial polarization decreases as excitation power is increased. The degree of polarization increases with time, reaching a peak of 70%. Fig. 3(d), shows the lifetime measurement corresponding to the white dashed lines in Fig. 3(a,c), showing an increase in the degree of polarization. We note that this effect was observed over multiple measurements and cooldowns. Similar results were also observed on an H-type stacked h-BN separated heterostructure, Denoted device 2 (Supplementary Fig. 6,7). We note that the H-type heterostructure shows an initial switch from negative to positive polarization but is then followed by an increase in polarization with time qualitatively consistent with Device 1.

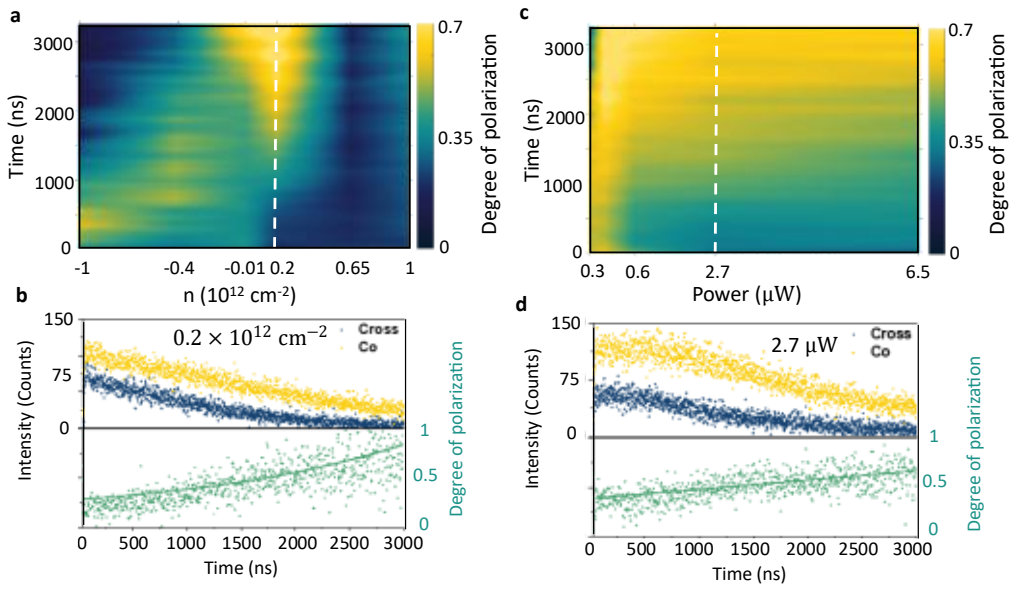

**FIG 3. Valley polarization dynamics**: (a) Degree of polarization as a function of time plotted for various doping levels at a fixed power of 0.74 μW. (b) The co and cross-polarized components of PL as a function of time (scan along the dashed line in (a)) show that while the cross-component intensity reaches zero by 3000 ns, the co-polarized intensity remains non-zero, leading to an increase in the degree of polarization. (c) Degree of polarization plotted as a function of time for different excitation powers while the sample is neutral. (d) Co and cross-polarized components of the lifetime plotted for 2.7 μW excitation power (scan along the dashed line).

To explain the observed polarization enhancement with time in hBN-separated heterostructures, we introduce a model for the valley polarization dynamics within the rate-equation framework. This model accounts for exciton diffusion, recombination, and intervalley exciton-exciton annihilation, as described by the following rate equations for the exciton density in the $+K$ and $-K$ valleys:

$$\frac{\partial n_+(t,r)}{\partial t} = D\nabla^2 n_+(t,\boldsymbol{r}) - \frac{n_+(t,r)}{\tau} - R_A n_+(t,\boldsymbol{r}) n_-(t,\boldsymbol{r}),$$
$$\frac{\partial n_-(t,r)}{\partial t} = D\nabla^2 n_-(t,\boldsymbol{r}) - \frac{n_-(t,r)}{\tau} - R_A n_-(t,\boldsymbol{r}) n_+(t,\boldsymbol{r}), \quad (1)$$

Here, $n_+$ and $n_-$ represent the exciton densities in the $+K$ and $-K$ valleys, respectively, and $\boldsymbol{r}$ is the in-plane position. The $R_A$ term describes an effective intervalley exciton-exciton annihilation channel, motivated by exciton-exciton annihilation models in TMDCs where exciton dynamics are commonly described using density-dependent Auger-type terms [33,34]. Following Ref. [35], we assume that the intervalley channel dominates in this system. Figure 4 shows simulations with an effective intervalley annihilation coefficient $R_A = 1.5 \times 10^{-4}\ \mathrm{cm^2/s}$, diffusion coefficient $D = 0.15\ \mathrm{cm^2/s}$ [36], and interlayer exciton lifetime $\tau = 2.0$ μs, consistent with the hBN-separated IX lifetime reported in Fig. 1(f). The intervalley scattering time was set to infinity in our model because previous reports of the intervalley scattering time for similar structures were on the order of 40 μs [37]. The initial exciton distribution is modeled as a Gaussian profile with a beam waist of 0.7 μm, matching the experimental laser spot diameter of 1.4 μm. The peak initial densities are set to $1.32 \times 10^{12}\ \mathrm{cm^{-2}}$ for the $+K$ valley and $0.77 \times 10^{12}\mathrm{cm^{-2}}$ for the $-K$ valley, based on experimental measurements. The total exciton number in each valley is obtained by integrating over position:

$$n_+(t) = \int n_+(r,t) d^2r$$
$$n_-(t) = \int n_-(r,t) d^2r. \quad (2)$$

The valley polarization is then computed as:

$$P(t) = \frac{n_+(t) - n_-(t)}{n_+(t) + n_-(t)}. \quad (3)$$

In Figs. 4(a) and 4(b), we present the exciton dynamics and polarization evolution. Figure 4(a) depicts the temporal evolution of the integrated exciton populations in the $+K$ and $-K$ valleys. Because the $-K$ valley starts with a lower initial population, the intervalley annihilation term gives it a larger effective per-particle nonlinear decay rate. As a result, the $-K$ population decreases more rapidly relative to its initial value, while the $+K$ population remains finite over the observed duration. This imbalance results in a time-dependent enhancement of valley polarization. Figure 4(b) shows the calculated degree of polarization, which exhibits a clear increase over time consistent with our experimental observation. In the absence of nonlinear intervalley annihilation, i.e., $R_A = 0$, Eq. (1) reduces to symmetric diffusion and recombination dynamics and would produce a constant polarization.

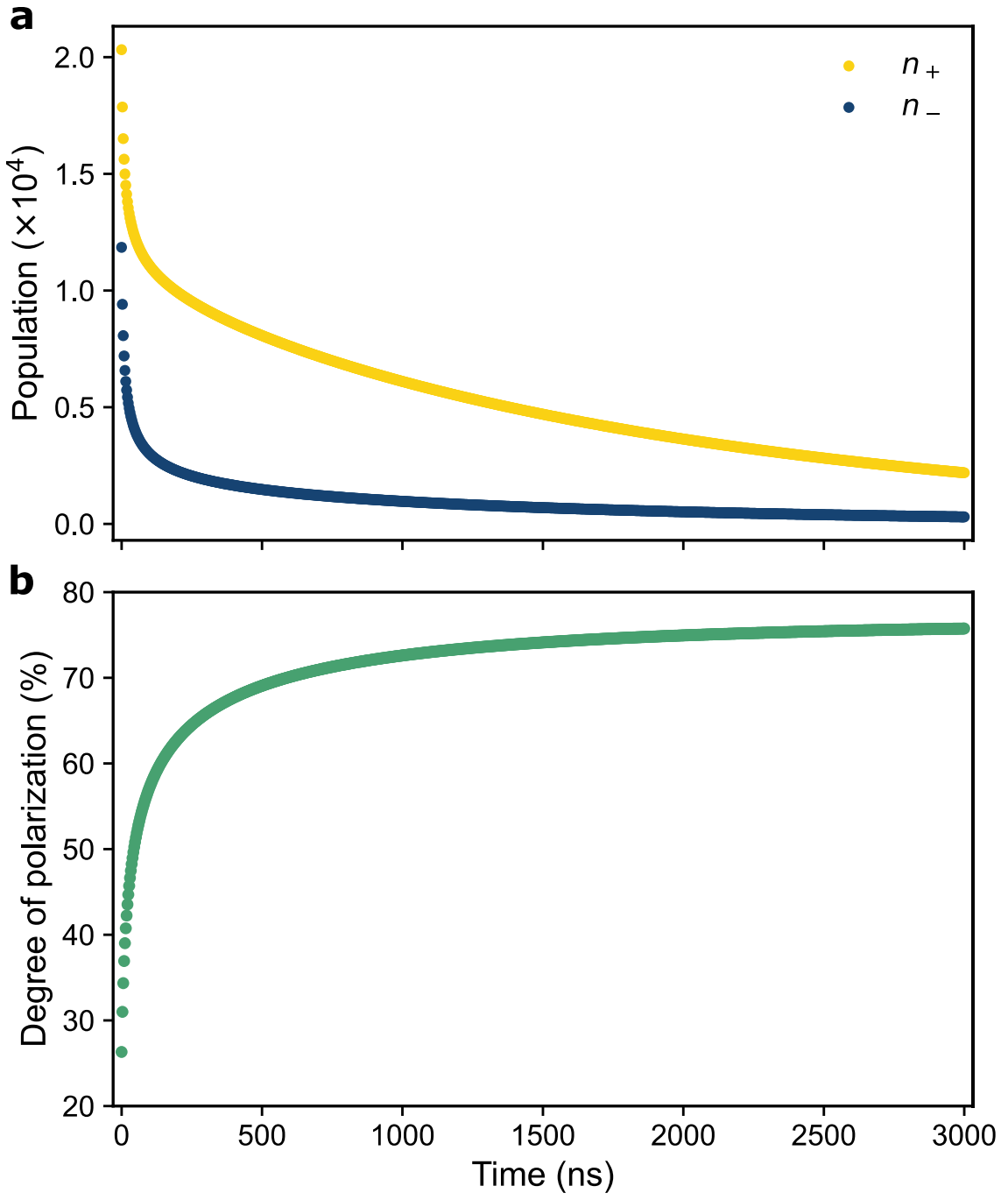

**FIG 4. Theoretical IX and polarization dynamics**. (a) Exciton density dynamics within valleys $+K$ ($n_+$) and $-K$ ($n_-$) (b) Degree of valley polarization as a function of time, highlighting the polarization amplification effect.

## IV. CONCLUSIONS

Our study reveals novel dynamics in the valley polarization of interlayer excitons within R-type heterostructures. These effects are attributable to an increase in the spatial separation between $MoSe_2/WSe_2$ monolayers. This pioneering approach not only extends the lifetime of these excitons beyond 2 µs, but also creates and increase in the degree of polarization over time. The enhancement in exciton stability within a specific valley results from nonlinear interactions, which are a key component in exciton-based logic devices. This valley polarization amplification marks a significant breakthrough for valleytronics.

## ACKNOWLEDGMENTS

This work is mainly supported by NSF Grant Nos. DMR-2003583, ECCS-2054572, and ECCS-2428575. V.P. acknowledges support from AFOSR Grant No. FA9550-22-1-0312. D.G.M. acknowledges support from the Gordon and Betty Moore Foundation's EPiQS Initiative, Grant GBMF9069. K.W. and T.T. acknowledge support from the CREST (JPMJCR24A5), JST and World Premier International Research Center Initiative (WPI), MEXT, Japan. B.J.L. and J.R.S. acknowledge additional support from Army Research Office under Grant no. W911NF-20-1-0215. J.R.S. acknowledges additional support from AFOSR Grant Nos. FA9550-20-1-0217, FA9550-21-1-0219, FA9550-22-1-0113, FA9550-22-1-0312, and FA9550-22-1-0220. B.J.L. acknowledges support from NSF Grant No. ECCS-2122462.